%% file: an-pn-IT2024.tex
\documentclass[journal, 11pt,one column,twosides]{IEEEtran} 

\usepackage[normalem]{ulem}

\newcommand{\stkout}[1]{\ifmmode\text{\sout{\ensuremath{#1}}}\else\sout{#1}\fi}

\usepackage[utf8]{inputenc} 
\usepackage[T1]{fontenc}
\usepackage{url}
\usepackage{ifthen}
\usepackage{cite}
\usepackage[cmex10]{amsmath} 
\usepackage{graphicx}
\usepackage{caption}
\usepackage{subcaption}
\usepackage{color}
\usepackage{epsfig}
\usepackage{amssymb}
\usepackage{amsmath}
\usepackage{amsthm}
\usepackage{soul}
\usepackage{latexsym}
\usepackage{setspace}
\usepackage{bbm}
\usepackage{arydshln}
\usepackage{textcomp}
\usepackage{amsfonts}
\usepackage{blindtext}
\usepackage{enumerate}
\usepackage{bbm}
\usepackage{upgreek}
\usepackage[bottom]{footmisc}
\usepackage{tikz}
\newcommand{\qeed}{\hfill $\blacksquare$}

\include{preamble}
\usepackage{secdot}
\allowdisplaybreaks
\sectiondot{subsection}
\sectiondot{subsubsection}
\usepackage{mleftright}

\usepackage{chngcntr}

\begin{document}

\newcommand\MC{{ \ - \!\!\circ\!\! - \ }}

\theoremstyle{theorem}
\newtheorem{theorem}{Theorem}
\newtheorem{corollary}[theorem]{Corollary}
\newtheorem{lemma}[theorem]{Lemma}
\newtheorem{proposition}[theorem]{Proposition}
\theoremstyle{definition}
\newtheorem{definition}{Definition}

\title{List Privacy Under Function Recoverability} 

\author{Ajaykrishnan Nageswaran and Prakash Narayan$^\dag$ }
\maketitle
{\renewcommand{\thefootnote}{}\vspace{-0.2cm}\footnotetext{
$^\dag$A. Nageswaran was with the Department of
Electrical and Computer Engineering and the Institute for Systems
Research, University of Maryland, College Park, MD 20742, USA, and is now a Research Scientist Resident 
at SandboxAQ, NY 10591, USA. 
E-mail: ajaykrishnan.n@gmail.com. P. Narayan is with the Department of
Electrical and Computer Engineering and the Institute for Systems
Research, University of Maryland, College Park, MD 20742, USA.
E-mail: prakash@umd.edu. This work was supported by the U.S.
National Science Foundation under Grant CCF 1527354. This paper was presented in
part at the $2019$ International Symposium on Information Theory~\cite{Nages19-2}.}}

\maketitle

\maketitle

\begin{abstract}
For a given function of user data, 
a querier must recover with at least a prescribed probability, the value of the function 
based on a user-provided query response. Subject to this requirement, the user forms the query response 
so as to minimize the likelihood of the querier guessing a list of prescribed size to which the data value belongs based 
on the query response. We obtain a general converse upper bound for maximum list privacy. This bound is shown to be 
tight for the case of a binary-valued function through an explicit achievability scheme that involves an add-noise query response.
\end{abstract}

\section{Introduction}
\label{sec:intro}

\ifx$\stackrel{\text{Noise}}{\leadsto}$\fi
A (legitimate) user's data is depicted by a finite-valued random variable (rv)
with known probability mass function (pmf). A querier wishes to compute a given function of the 
data from a query response provided by the user that is a suitably randomized version of the data. The user 
fashions the query response so as to enable the querier to recover from it the function value with a 
prescribed accuracy while maximizing ``list privacy'' of the data. List privacy entails minimizing the likelihood of 
the querier guessing a list of prescribed size to which the data value belongs based on the query response.

For example, in a movie-streaming application where a user requests a particular movie (data), the identity of 
that movie is protected, say, by end-to-end encryption. However, the server application could still be interested 
in the genre of movies that are being streamed (list), which the user wishes to be kept private. 
The movie requested plays the role of user data, and the server application compiles an estimated list of movies 
sharing the same genre as the movie. This list should not include the user's movie choice. In another instance, 
a list of call detail records (list) of multiple users is collected, for example, through a phone records sweep. Then 
information concerning an individual user (data) needs to be protected, but could be compromised upon whittling 
down the list using other available information such as phone area codes. In this example, the individual 
user information constitutes user data, and the call detail records form the list in which the user data should 
not be contained.

Our contributions, which represent a continuation of our earlier work concerning {\it data privacy}~\cite{Nages18},~\cite{Nages19}, 
are as follows. In our formulation, the user forms a 
query response from which the querier must recover the function value with probability at least $\rho$, 
$0\leq\rho\leq 1$. Under this requirement, we introduce the notion of \textit{list privacy} 
that is a generalization of data privacy in~\cite{Nages18},~\cite{Nages19}. List privacy is measured by the probability that 
the user data is not contained in a list of prescribed size formed by the querier based on the query response. In fact, list 
privacy is a more stringent notion than a certain level of data privacy; and, as will be seen below ensuring list privacy for the 
user guarantees at least the same level of data privacy. We derive a general converse upper bound for maximum list privacy. 
For the special case of a binary-valued function, the mentioned 
upper bound is shown to be tight through an achievability scheme based on an explicit query response. 
A key theoretical insight is described in the second paragraph of Section~\ref{sec:list_privacy} below.

A notion of list privacy -- in an asymptotic rate-distortion framework -- has been considered earlier in a cryptographic context,
for instance, in~\cite{schieler16}. In~\cite{schieler16}, a new concept of information-theoretic secrecy is studied in the Shannon 
cipher system setting where the eavesdropper produces a list of reconstruction sequences, and secrecy is measured by the 
minimum distortion over the entire list. Of a different nature, our approach is in the spirit of prior works~\cite{CalmonFawaz12},~\cite{Sankar13},~\cite{ MakhdoumiFawaz13},~\cite{Huang17},~\cite{Wang21} 
that deal with information leakage of a user's private data with associated nonprivate correlated data. A 
randomized version of the nonprivate data is released publicly under a constraint on the expected distortion 
between the nonprivate and public data. The public data is designed such that it minimizes 
information leakage, which is measured by the mutual information between public and private data, under
 said distortion constraint. These works are based on principles of rate distortion theory.

In a related data privacy model in~\cite{Asoodeh18},~\cite{Diaz20}, maximum a posteriori (MAP) estimates of private and 
nonprivate data, that are digital, are formed on the basis of randomized public versions of the latter. The private, nonprivate 
and public data are assumed to form a Markov chain. Under a constraint on the probability of estimating 
correctly the private data, mechanisms are sought for said randomization so as to maximize correct MAP 
estimation of the nonprivate data.

In data privacy, an important movement that has commanded dominant attention over the years is differential privacy, introduced 
in~\cite{Dwork06},~\cite{DworkSmith06} and explored further in~\cite{McSherry07},~\cite{Bassily13},~\cite{Kasi14},~\cite{Mironov17} among 
others. Consider a data vector that represents multiple users' data. The notion of 
differential privacy stipulates that altering a data vector slightly leads only to a near-imperceptible 
change in the corresponding probability distribution of the output of the data release mechanism, which 
is a randomized function of the data vector. Upon imposing a differential privacy constraint, there exists a large body of 
work that seeks to maximize function recoverability by minimizing a discrepancy cost between function value and 
randomized query response, a representative sample is~\cite{Hardt10},~\cite{Wasserman10},~\cite{Geng16},~\cite{Geng20}. 
\textit{In contrast, our work maximizes privacy 
under a constraint on recoverability}. Also, see~\cite{Wasserman10} for data privacy methods other than differential privacy.

Our model for $\rho$-recoverable function computation with associated list privacy is 
described in Section~\ref{sec:prelim}. List privacy is characterized in Section~\ref{sec:list_privacy} and 
Section~\ref{sec:list-priv-disc} contains a closing discussion.

\section{Preliminaries}
\label{sec:prelim}

Let a (legitimate) user's data be represented by a rv $X$ taking values in a finite 
set $\cX$ with $|\cX|=r\geq 2$, say, and of known pmf $P_X$ with $P_X\left(x\right)>0,$ $x\in\cX.$  
We shall consider throughout a given mapping $f:\cX\rightarrow\cZ=\{0,1,\ldots,k-1\},$ $2\leq k\leq r.$ 
Let $f^{-1}$ denote the inverse image of $f$, with 
$f^{-1}(i)\triangleq\left\{x\in\cX:f(x)=i\right\}, \ i\in\cZ$.
For a realization $X=x$ in $\cX,$ a querier -- who does not know $x$ -- wishes to compute $f\left(x\right)$ 
from a \textit{query response} (QR) $F\left(x\right)$ provided by the user, where $F\left(x\right)$ is a rv 
with values in $\cZ.$ A QR must satisfy the following recoverability condition. 

\vspace{0.10cm}

\begin{definition}
\label{def:rho-recov}
Given $0\leq\rho\leq 1,$ a QR $F\left(X\right)$ is 
$\rho$\textit{-recoverable} if 
\begin{equation}
\label{eq:recov1}
P\left(F\left(X\right)=f\left(x\right)\big|X=x\right)\geq 
\rho,\hspace{4mm}x\in\cX.
\end{equation} 
Condition~\eqref{eq:recov1} can be written equivalently 
in terms of a stochastic matrix $W:\cX\rightarrow\cZ$ with the 
requirement 
\begin{equation}
\label{eq:recov2}
W\big(f\left(x\right)|x\big)\geq \rho,\hspace{4mm}x\in\cX
\end{equation}

\noindent which, too, will constitute a $\rho$\textit{-recoverable} QR.
Such a $\rho$-recoverable $F\left(X\right)$ or $W$ will be termed $\rho$-QR. Note that $\rho$-recoverability 
in~\eqref{eq:recov1},~\eqref{eq:recov2} does not depend on $P_X$.
\end{definition}
\vspace{0.1cm}
In order to give additional teeth to our notion of privacy, we assume conservatively 
that the querier knows the pmf $P_X$ and the $\rho$-QR $F$ or $W$.

\begin{definition}
\label{def:list_privacy}
Let $\mathcal{L}_l$ be the set of all $l$-sized subsets of $\cX$, $1\leq l<r$. 
For a given $l$, $1\leq l< r$, the ($l$-)\textit{list privacy} of a $\rho$-QR $F(X)$ or 
$W$ satisfying~\eqref{eq:recov1} or~\eqref{eq:recov2} is
\begin{equation}
\label{eq:list_priv}
\pi^{(l)}_{\rho}(F) = \pi^{(l)}_{\rho}(W) \triangleq \min_{g} \ P\left(X\notin g\left(F(X)\right)\right),
\end{equation}
\noindent where the minimum is over all estimators $g:\cZ\rightarrow\mathcal{L}_l$. The 
minimum in~\eqref{eq:list_priv} is attained by a MAP estimator

\begin{equation}
\label{eq:map_list_privacy}
g_{MAP(W)}^{(l)}(i) = \arg\hspace{0.1cm}\max_{L\in\cL_l}\hspace{1.5mm}
P\left(X\in L,F(X)=i\right)\\=\arg\hspace{0.1cm}\max_{L\in\cL_l}\hspace{1.5mm}
\sum\limits_{x\in L}P_X\left(x\right)W\left(i|x\right),\hspace{4mm}i\in\cZ.
\end{equation}
\noindent Ties in~\eqref{eq:map_list_privacy} are broken arbitrarily. For $0\leq\rho\leq 1$, the maximum list privacy that can be 
attained by a $\rho$-QR is termed \textit{list $\rho$-privacy} and denoted by 
$\pi^{(l)}(\rho)$, i.e.,
\begin{equation*}
\pi^{(l)}\left(\rho\right)\triangleq \max_{\substack{W : W \left( f \left(x\right) | x 
\right)\geq\rho \\ x\in\cX}}  \pi_\rho^{(l)}\left(W\right).
\end{equation*}
\end{definition}
\vspace{.3cm}

Our objective is to characterize list $\rho$-privacy, and identify 
\textit{explicitly} $\rho$-QRs that achieve them, $0\leq\rho\leq 1$. This objective is met below in part for $\pi^{(l)}(\rho)$.
\vspace{0.2cm}\\

\noindent\textit{Remark}: Our earlier works~\cite{Nages18},~\cite{Nages19} were concerned primarily with privacy for the 
data $X$. The associated concept of $\rho$-\textit{privacy} $\pi(\rho)$ for $X$ is a special 
case of list $\rho$-privacy $\pi^{(l)}(\rho)$ for list size $l=1$. Clearly, list 
privacy is a more stringent notion than (data) privacy since $\pi^{(l)}(\rho)\leq\pi(\rho)$. Also, for $1\leq l'<l<r$, $\pi^{(l)}(\rho)\leq \pi^{(l')}(\rho)$.

%%%%%%%%%%%%%%%%%%%%%%%%%%%%%%%%%%%%%%%%%%%%%%%%%%%%%%%%%%%%%%%%%%%%%%%%%%%%%%%%%%%%%%%%%%%%%%

\section{List $\rho$-Privacy}
\label{sec:list_privacy}

\vspace{0.2cm}

Our characterization of list $\rho$-privacy $\pi^{(l)}(\rho)$, $0\leq\rho\leq 1$, 
is partial. We provide an upper bound for $\pi^{(l)}(\rho)$, and an achievability proof for the special case of a 
binary-valued mapping $f:\cX\rightarrow\cZ=\{0,1\}$. A general achievability proof is not yet in hand.

Our first main result is a general upper bound for $\pi^{(l)}(\rho), \ 0\leq\rho\leq 1$. This upper bound below is guided by 
the following heuristics. When $\rho=0$, the $\rho$-QR $F(X)$ gives no actionable information to the querier. Therefore, the 
querier's best list estimate consists of the $l$ largest $P_X$-probability elements of the data $X$ in $\cX$, and the corresponding 
list estimation error probability is $\pi^{(l)}(0)$. The other extreme of $\rho=1$ corresponds to the querier having a full knowledge 
of $f(X)$. Then, for $f(X)=i$ in $\cZ$, the best list estimate for the querier will contain the largest $\min\{l,|f^{-1}(i)|\}$ $P_X$-probability 
elements in the subset $f^{-1}(i)$ of $\cX$. For any intermediate value of $\rho$, the querier's optimal choice will be a suitable 
combination of these two extremes. In other words, the querier's best list estimate will consist of a mix of the highest $P_X$-probability 
elements in $\cX$ and the highest $P_X$-probability elements in $f^{-1}(i)$, for $F(X)=i$ in $\cZ$; and as $\rho$ increases, the 
relative size of the latter will increase while the former diminishes.

\noindent Specifically, for any $A\subseteq \cX$, we denote by $\left[A\right]_t$
the set of $t$ largest $P_X$-probability elements in $A$, $0\leq t\leq |A|\leq r$; in particular, $\left[A\right]_0=\emptyset$ 
and $L_t^*\triangleq\left[\cX\right]_t$ 
is the set of $t$ largest $P_X$-probability elements in $\cX$. Define

\begin{equation}
\label{eq:l_rho}
\Lambda_{\rho}\triangleq\arg \ \max_{\substack{\Lambda \subset \cX: \\ 0\leq |\Lambda|\leq l}} 
\left[ P_X(\Lambda) + \rho\sum\limits_{i\in\cZ} P_X\left(\left[f^{-1}(i)\setminus 
\Lambda\right]_{\min\{l-|\Lambda|,|f^{-1}(i)\setminus \Lambda|\}} \right)\right], \ \ \ 0\leq\rho\leq 1.
\end{equation}

\begin{theorem}
\label{thm:list_priv_ub}
\begin{enumerate}[(i)]
\item For $0\leq\rho\leq 1/k$,
\begin{equation}
\label{eq:priv-at-zero}
\pi^{(l)}(\rho)=1-P_X\left(L_l^*\right).
\end{equation}
\item For $\rho=1$,
\begin{equation}
\label{eq:priv-at-one}
\pi^{(l)}(1)=1-\sum\limits_{i\in\cZ} \ P_X\left(\left[f^{-1}(i)\right]_{\min\{l,|f^{-1}(i)|\}}\right).
\end{equation}
\item For $0\leq\rho\leq 1$, $\pi^{(l)}(\rho)$ is bounded above according to 
\begin{equation}
\label{eq:list_privacy_ub}
\pi^{(l)}(\rho) \leq \pi_u^{(l)}(\rho) \triangleq 1 - 
\left[ P_X(\Lambda_{\rho}) + \rho\sum\limits_{i\in\cZ} P_X\left(\left[f^{-1}(i)\setminus 
\Lambda_{\rho}\right]_{\min\{l-|\Lambda_{\rho}|,|f^{-1}(i)\setminus \Lambda_{\rho}|\}} \right)\right].
\end{equation}
\end{enumerate}
\end{theorem}
\vspace{0.1cm}

\noindent\textit{Remarks}:
\vspace{0.1cm}
\begin{enumerate}[(i)]
\item In the (uninteresting) regime of low $\rho$, specifically $0\leq\rho\leq 1/k$, Theorem~\ref{thm:list_priv_ub}(i) 
shows that the $\rho$-QR $F(X)$ is useless for the querier whose best estimate of a list is $L_l^*$, i.e., 
the list of $l$-largest $P_X$-probability elements in $\cX$.
\item For $\rho= 1$, when the querier has 
 full knowledge of $f(X)$, say $f(X)=i$ in $\cZ$, the querier's best list estimate $L_l$ will 
 contain $\left[f^{-1}(i)\right]_{\min\{l,|f^{-1}(i)|\}}$. Theorem~\ref{thm:list_priv_ub}(ii) characterizes the corresponding 
 $\rho$-privacy $\pi^{(l)}(1)$. 
 \item  As described above, the form of $\pi_u^{(l)}(\rho)$ in~\eqref{eq:list_privacy_ub},~\eqref{eq:l_rho} 
affords the following interpretation for the querier's best list estimate: For any intermediate value of $\rho$, the expression 
in~\eqref{eq:list_privacy_ub} 
captures a tradeoff between the extremes in Remarks (i), (ii).
\item The maximizing $\Lambda=\Lambda_{\rho}$ in~\eqref{eq:l_rho} need not be unique.
\item In the right-side of~\eqref{eq:list_privacy_ub}, $\pi_u^{(l)}(\rho)$ is piecewise affine in $\rho\in [0,1]$. See, for example, 
Fig.~\ref{fig:list-priv}. 

\item For $l=1$, 
\[\pi_u^{(l)}(\rho)=1-\max\left\{\max_{x\in\cX}\hspace{0.1cm}P_X(x),\rho\sum\limits_{i\in\cZ}
\max_{x\in f^{-1}(i)}\hspace{0.1cm}P_X(x)\right\}\]
\noindent which, by~{\cite[Theorem $2$]{Nages19}} equals $\rho$-privacy $\pi(\rho)$. 
Thus, Theorem~\ref{thm:list_priv_ub} is tight for $l=1$.
\end{enumerate}

\vspace{0.2cm}
\begin{figure}[h!]
\centering
\includegraphics{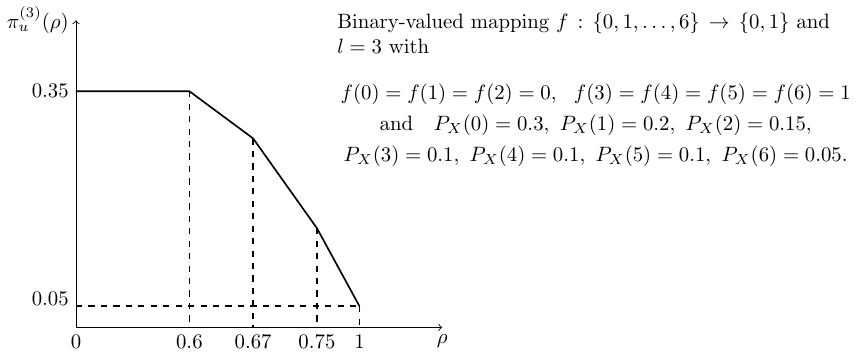}
\caption{$\pi_u^{(3)}(\rho)$ vs. $\rho$}
\label{fig:list-priv}
\end{figure}
\ifx
\begin{figure}[h!]
\centering
\begin{tikzpicture}[scale=0.85]
\draw[->] (0,0) -- (7.75,0) node[anchor=north] {$\rho$};
\draw	(0,0) node[anchor=north] {0}
		(2.4,0) node[anchor=north] {$0.6$}
		(3.75,0) node[anchor=north] {$0.67$}
		(5.1,0) node[anchor=north] {$0.75$}
		(6,0) node[anchor=north] {1};
\draw	(0,0.6) node[anchor=east]{$0.05$}
		(0,5) node[anchor=east] {$0.35$};
  \node[text width=8cm] at (10,6) {Binary-valued mapping $f:\{0,1,\ldots,6\}\rightarrow\{0,1\}$ and $l=3$ with 
\begin{gather*}
    f(0)=f(1)=f(2)=0, \ \   f(3)=f(4)=f(5)=f(6)=1 \\ \text{and} \ \ \ P_X(0)=0.3, \ P_X(1)=0.2, \ P_X(2)=0.15, \\ 
    P_X(3)=0.1, \ P_X(4)=0.1, \ P_X(5)=0.1, \ P_X(6)=0.05. 
\end{gather*}};

\draw[->] (0,0) -- (0,6.5) node[anchor=east] {$\pi_u^{(3)}(\rho)$};

\draw[thick] (0,5) -- (2.4,5) -- (3.75,4) -- (5.1,2.1) -- (6,0.45);
\draw[thick,dashed] (2.4,5) -- (2.4,0);
\draw[thick,dashed] (3.75,4) -- (3.75,0);
\draw[thick,dashed] (5.1,2.1) -- (5.1,0);
\draw[thick,dashed] (6,0.45) -- (6,0);
\draw[thick,dashed] (0,0.45) -- (6,0.45);
\end{tikzpicture}
\caption{$\pi_u^{(3)}(\rho)$ vs. $\rho$}
\label{fig:list-priv}
\end{figure}
\fi

For the special case of binary-valued functions $f:\cX\rightarrow\cZ=\{0,1\}$, the upper bound in 
Theorem~\ref{thm:list_priv_ub}(iii) is tight. This is shown by identifying a suitable $\rho$-QR that achieves said bound.

A special form of $\rho$-QR, termed add-noise $\rho$-QR in{\cite[Definition $2$]{Nages19}}, will be 
pertinent. An {\em add-noise} $\rho$-QR $F(X)$ is of the form
\begin{equation}
    \label{eq:add-noise-QR}
    F(X)=f(X)+N \ \ \mod k
\end{equation}
\noindent where $N$ is a $\cZ$-valued rv that satisfies
\begin{equation}
 \label{eq:mark_cond}
 N\MC f(X)\MC X
\end{equation}
\noindent whereby readily implying 
\begin{equation}
 \label{eq:mark_cond_imply}
 X\MC f(X)\MC F(X).
\end{equation}

As shown in{\cite[Lemma $1$]{Nages19}}, such an add-noise privacy mechanism corresponds to 
a stochastic matrix $W:\cX\rightarrow\cZ$ with identical rows for all $x\in f^{-1}(i), \ i\in\cZ$.

\begin{theorem}
\label{thm:achiev}
For $f:\cX\rightarrow\cZ=\{0,1\}$, it holds that list $\rho$-privacy is
\begin{equation}
\label{eq:list-priv-binary}
\pi^{(l)}(\rho) = \pi_u^{(l)}(\rho)=1 -  
\left[ P_X(\Lambda_{\rho}) + \rho\sum\limits_{i\in\{0,1\}} P_X\left(\left[f^{-1}(i)\setminus 
\Lambda_{\rho}\right]_{l-|\Lambda_{\rho}|} \right)\right], 
 \ \ \ 0\leq\rho\leq 1,
 \end{equation}
 \noindent and is achieved by an add-noise $\rho$-QR.
\end{theorem}
\vspace{0.1cm}
\noindent\textit{Remark}: As noted below in the proof of Theorem~\ref{thm:achiev}, the minimum 
in~\eqref{eq:list_privacy_ub} is attained by $l-|\Lambda_{\rho}|$.
\vspace{0.2cm}\\
\noindent\textit{Example}: This example illustrates the remark at the end of Section~\ref{sec:prelim}. 
Let $X$ be a $\{0,1,2,3\}$-valued rv with uniform pmf $P_X$. Consider the mapping $f:\{0,1,2,3\}\rightarrow\{0,1\}$ with 
\[
f(0)=f(1)=0, \ \ \ \ \ f(2)=f(3)=1.
\]

\noindent By Remark (vi) after Theorem~\ref{thm:list_priv_ub} above,
\[\pi(\rho)=\pi^{(1)}(\rho)=1-\max\{0.25,0.5\rho\}, \ \ \ 0\leq\rho\leq 1.\]
\noindent Next, in~\eqref{eq:l_rho}, we have
\begin{equation*}
\Lambda_{\rho}=\arg \ \max_{\substack{\Lambda \subset \cX: \\ 0\leq |\Lambda|\leq l}} 
\left[ 0.25\left|\Lambda\right|+ 0.25\rho\sum\limits_{i\in\{0,1\}} {\min\{l-|\Lambda|,|f^{-1}(i)\setminus \Lambda|\}} \right], \ \ \ 0\leq\rho\leq 1,
\end{equation*}
\noindent and so by Theorem~\ref{thm:achiev},
\[\pi^{(2)}(\rho)=1-\max\{0.5,\rho\}, \ \ \ 0\leq\rho\leq 1,\]
\noindent and
\[\pi^{(3)}(\rho)=1-\max\{0.75,0.5+0.5\rho\}, \ \ \ 0\leq\rho\leq 1.\]
\noindent Thus, for $0\leq\rho<1$,
\[\pi^{(1)}(\rho)>\pi^{(2)}(\rho)>\pi^{(3)}(\rho)>\pi^{(4)}(\rho)=0,\]
\noindent where the last equality is obvious.  \qeed

\vspace{0.1cm}
The following technical 
lemma, whose proof is relegated to Appendix~\ref{app:lemma-pf}, states properties of $\Lambda_{\rho}$ and $\pi_u^{(l)}(\rho)$ 
in~\eqref{eq:l_rho},~\eqref{eq:list_privacy_ub} of pertinence to the  proof of Theorem~\ref{thm:achiev}.
\begin{lemma}
\label{lem:list_priv_ub}
$\big($Properties of $\Lambda_{\rho}$ and $\pi_u^{(l)}(\rho)$ in~\eqref{eq:l_rho},
~\eqref{eq:list_privacy_ub},~\eqref{eq:list-priv-binary}$\big)$

\begin{enumerate}[(i)]
\item $\Lambda_{\rho}=\bigcup\limits_{i\in\cZ}\left(\Lambda_{\rho}\cap f^{-1}(i)\right)=
\bigcup\limits_{i\in\cZ}\left[f^{-1}(i)\right]_{\left\lvert \Lambda_{\rho}\cap f^{-1}(i)\right\rvert}$, i.e., 
$\Lambda_{\rho}\cap f^{-1}(i)$ consists of $\left\lvert \Lambda_{\rho}\cap 
f^{-1}(i)\right\rvert$-largest 
$P_X$-probability elements in $f^{-1}(i), \ i\in \cZ$.
\item For $f:\cX\rightarrow\cZ=\{0,1\}$, $\Lambda_{\rho}$ satisfies
\begin{equation}
    \label{eq:l-rho-not-empty}
    l-|\Lambda_{\rho}|\leq |f^{-1}(i)\setminus \Lambda_{\rho}|, \ \ \ i\in\{0,1\}.
\end{equation}
    \item $\pi_u^{(l)}(\rho)$ in~\eqref{eq:l_rho},~\eqref{eq:list_privacy_ub} is piecewise affine, continuous and nonincreasing 
 in $\rho\in [0,1]$, with 
 \begin{equation}
 \label{eq:list_priv_extremes}
 \pi_u^{(l)}(0) = 1-P_X\left(L_l^*\right),\hspace{0.2cm}\pi_u^{(l)}(1) = 
 1-\sum\limits_{i\in\cZ}P_X\left(\left[f^{-1}(i)\right]_{\min\{l,|f^{-1}(i)|\}}\right).
 \end{equation}
 \item $|\Lambda_{\rho}|$ is nonincreasing in $\rho, \ 0\leq\rho\leq 1$, with $|\Lambda_0|=l$ and 
 $0\leq|\Lambda_1|\leq \max\{0,l-\min\limits_{i\in\cZ} \ |f^{-1}(i)|\}$. 
 \item $|\Lambda_{\rho}|$ is piecewise constant in $0\leq\rho\leq 1$ with $|\Lambda_{\rho}|=l-j$ for $\rho\in [\rho_j,
 \rho_{j+1}]$, $0\leq j\leq l$, where\\ $\rho_0=0<\rho_1\leq \ldots\leq\rho_l\leq\rho_{l+1}=1;$ 
 and $\rho_j$, $j=1,\ldots,l$, is the solution to
 \begin{multline}
 \label{eq:lemma-2v}
\max_{\substack{\Lambda \subset \cX: \\ |\Lambda|=l-j+1}}\left[
 P_X(\Lambda) + \rho\sum\limits_{i\in\cZ} P_X\left(\left[f^{-1}(i)
 \setminus \Lambda\right]_{\min\{j-1,|f^{-1}(i)\setminus \Lambda|\}}\right) \right]\\
=  \max_{\substack{\Lambda' \subset \cX: \\ |\Lambda'|=l-j}}\left[
 P_X(\Lambda') + \rho\sum\limits_{i\in\cZ} P_X\left(\left[f^{-1}(i)
 \setminus \Lambda'\right]_{\min\{j,|f^{-1}(i)\setminus \Lambda'|\}}\right) \right].
 \end{multline}
\end{enumerate}
\end{lemma}

\vspace{0.3cm}
\noindent\textit{Proof of Theorem~\ref{thm:list_priv_ub}}: 
\vspace{0.2cm}

(i) For every $\rho$-QR $W:\cX\rightarrow\cZ$ in~\eqref{eq:recov2},
\begin{align}
1-\pi_{\rho}^{(l)}(W)&=P\left(X \in g_{MAP(W)}^{(l)}\left(F(X)\right)\right)\nonumber\\
&=\sum\limits_{i\in\cZ} P\left(X \in g_{MAP(W)}^{(l)}\left(F(X)\right),F(X)=i\right)\nonumber\\
&=\sum\limits_{i\in\cZ}\hspace{0.04cm} \max_{L\in\cL_l}\hspace{0.04cm} P\left(X \in L,F(X)=i\right)\nonumber\\
&=\sum\limits_{i\in\cZ}\hspace{0.04cm} \max_{L\in\cL_l}\hspace{0.04cm}\sum\limits_{x\in L} P_X(x) W(i|x)\label{eq:loplop}\\
&\geq \sum\limits_{i\in\cZ}\hspace{0.04cm} \sum\limits_{x\in L_l^*} P_X(x) W(i|x) \nonumber\\
&= \sum\limits_{x\in L_l^*}\hspace{0.04cm}P_X(x)\sum\limits_{i\in\cZ}   W(i|x)=P_X\left(L_l^*\right),\nonumber
\end{align}
\noindent where the inequality above is by choosing $L=L_l^*$ in~\eqref{eq:loplop}. Therefore, 
\begin{equation}
\label{eq:priv-at-zero-ub}
\pi^{(l)}(\rho)\leq 1-P_X\left(L_l^*\right), \ \ \ 0\leq\rho\leq 1.
\end{equation}
\noindent The upper bound in~\eqref{eq:priv-at-zero-ub} can be achieved by a $\rho$-QR $W_0:\cX\rightarrow\cZ$, $0\leq\rho\leq 1/k$, given by
\[
W_0(i|x)=\frac{1}{k}, \ \ \ i\in\cZ, \ x\in\cX.
\]
Using~\eqref{eq:loplop},
\[ \pi_{\rho}^{(l)}\left(W_0\right)=1-\sum\limits_{i\in\cZ}\hspace{0.04cm} \max_{L\in\cL_l}\hspace{0.04cm}
\sum\limits_{x\in L} P_X(x) W(i|x)=1-\sum\limits_{i\in\cZ}\hspace{0.04cm} \max_{L\in\cL_l}\hspace{0.04cm}
\sum\limits_{x\in L} P_X(x) \frac{1}{k}=1-P_X\left(L_l^*\right),\] which along with~\eqref{eq:priv-at-zero-ub} yields~\eqref{eq:priv-at-zero}.
\vspace{0.2cm}

(ii) For $\rho=1$, the only feasible $\rho$-QR is $W_1:\cX\rightarrow\cZ$ given by
\[
W_1(i|x)=\begin{cases}
    1,\ \ \ \ &x\in f^{-1}(i), \ i\in\cZ\\
    0, \ \ \ &\text{otherwise}.
\end{cases}
\]
\noindent Then
\begin{align*}
    1-\pi(1)&=1-\pi_{1}^{(l)}(W_1)\\
    &=\sum\limits_{i\in\cZ}\hspace{0.04cm} \max_{L\in\cL_l}\hspace{0.04cm}\sum\limits_{x\in L} P_X(x) W_1(i|x), \ \ \ \ \ \text{by~\eqref{eq:loplop}}\\
    &=\sum\limits_{i\in\cZ}\hspace{0.04cm} \max_{L\in\cL_l}\hspace{0.04cm}\sum\limits_{x\in L} P_X(x) \mathbbm{1}(i=f(x))\\
    &=\sum\limits_{i\in\cZ} \ P_X\left(\left[f^{-1}(i)\right]_{\min\{l,|f^{-1}(i)|\}}\right),
\end{align*}
\noindent which gives~\eqref{eq:priv-at-one}.
\vspace{0.2cm}

(iii) By~\eqref{eq:loplop}, for every $\rho$-QR $W:\cX\rightarrow\cZ$ in~\eqref{eq:recov2}, 
\begin{align}
1-\pi_{\rho}^{(l)}(W)&=\sum\limits_{i\in\cZ}\hspace{0.04cm} \max_{L\in\cL_l}\hspace{0.04cm}\sum\limits_{x\in L} P_X(x) W(i|x)\label{eq:sec-summ}\\
&\geq \sum\limits_{i\in\cZ}\hspace{0.1cm}\max_{\substack{\Lambda \subset \cX: \\ 0\leq |\Lambda|\leq l}} \ 
\sum\limits_{\substack{x\in \Lambda\cup\left[f^{-1}(i)\setminus \Lambda\right]_{\min\{l-|\Lambda|,|f^{-1}
(i)\setminus \Lambda|\}}}} P_X(x) W(i|x)\nonumber\\
&=\sum\limits_{i\in\cZ}\hspace{0.12cm} \max_{\substack{\Lambda \subset \cX: \\ 0\leq |\Lambda|\leq l}}\hspace{0.1cm}
\bigg[\sum\limits_{x\in \Lambda} P_X(x) W(i|x)+ 
\sum\limits_{x\in \left[f^{-1}(i)\setminus \Lambda\right]_{\min\{l-|\Lambda|,|f^{-1}(i)\setminus \Lambda|\}}} P_X(x) W(i|x)
\bigg]\nonumber\\
&\geq\max_{\substack{\Lambda \subset \cX: \\ 0\leq |\Lambda|\leq l}}\sum\limits_{i\in\cZ}\hspace{0.1cm}
\bigg[\sum\limits_{x\in \Lambda} P_X(x) W(i|x)+ 
\sum\limits_{x\in \left[f^{-1}(i)\setminus \Lambda\right]_{\min\{l-|\Lambda|,|f^{-1}(i)\setminus \Lambda|\}}} P_X(x) W(i|x)\bigg]\nonumber\\
&\geq \max_{\substack{\Lambda \subset \cX: \\ 0\leq|\Lambda|\leq l}}\left[
P_X(\Lambda) + \rho\sum\limits_{i\in\cZ} P_X\left(\left[f^{-1}(i)\setminus \Lambda\right]_{\min\{l-|\Lambda|,|f^{-1}(i)\setminus 
\Lambda|\}}\right) \right]\label{eq:bound_slack}
\end{align}

\noindent which yields~\eqref{eq:list_privacy_ub}. \qeed
\vspace{0.3cm}

\textit{ Proof of Theorem~\ref{thm:achiev}}:
\vspace{0.2cm}

For $f:\cX\rightarrow\{0,1\}$, we first note by Lemma~\ref{lem:list_priv_ub}(ii) that $\pi_{u}^{(l)}(\rho)$ 
in~\eqref{eq:list_privacy_ub} simplifies to the form of
 $\pi_{u}^{(l)}(\rho)$ in the right-side of~\eqref{eq:list-priv-binary}.

Next, we present a $\rho$-QR that achieves the right-side of~\eqref{eq:list-priv-binary}. In particular, this $\rho$-QR 
will involve $0<\rho_1<1$ where, by Lemma~\ref{lem:list_priv_ub}(v), with $j=1$ in~\eqref{eq:lemma-2v}, $\rho=\rho_1$ is the solution to 
\[
P_X(L_l^*)=\max_{\substack{\Lambda' \subset \cX: \\ |\Lambda'|=l-1}}\left[
 P_X(\Lambda') + \rho\sum\limits_{i\in\cZ} P_X\left(\left[f^{-1}(i)
 \setminus \Lambda'\right]_1\right) \right],
\]
\noindent observing that $\Lambda=L_l^*$ is the maximizer in the left-side of~\eqref{eq:lemma-2v}. Furthermore, the 
right-side of~\eqref{eq:list-priv-binary} simplifies for $0\leq\rho\leq\rho_1$. Specifically, by Lemma~\ref{lem:list_priv_ub}(v), 
$|\Lambda_{\rho}|=l, \ 0\leq\rho\leq\rho_1$, and by the observation above $\Lambda_{\rho}=L_l^*, 0\leq\rho\leq\rho_1$. 
Hence, by the right-side of~\eqref{eq:list-priv-binary},

\begin{equation}
\label{eq:list_priv_proof}
\pi_u^{(l)}(\rho) = 1-P_X\left(L_l^*\right), \ \ \ \ 0\leq\rho\leq\rho_1.
 \end{equation}
\noindent Clearly, by Theorem~\ref{thm:list_priv_ub}(i), $\rho_1\geq1/k$. Consider the add-noise 
$\rho$-QR~\eqref{eq:add-noise-QR}-\eqref{eq:mark_cond_imply}
\[
F(X)=f(X)+N \ \mod 2
\]

\noindent with
\begin{equation*}
    P\left(N=0 \big| f(X)=i\right) = \max\{\rho,\rho_1\}, \ \ \ i\in\{0,1\}
\end{equation*}\noindent where, in this instance, $N$ is independent of $f(X)$. The $\rho$-QR above corresponds to 
$W_o:\cX\rightarrow \{0,1\}$ given by
\begin{equation}
    \label{eq:achiev-scheme}
W_o(i|x) = \begin{cases}
\max\{\rho,\rho_1\},\ \ &i=f(x)\\
1-\max\{\rho,\rho_1\},\ \ &i\neq f(x), \ \ x\in\cX, \ \ i\in\{0,1\}.
\end{cases}
\end{equation} 
The list privacy afforded by $W_o$ is
\begin{align}
\pi_{\rho}^{(l)}\left(W_o\right)&=1-P\left(X \in g_{MAP\left(W_o\right)}^{(l)}\left(F(X)\right)\right)\nonumber\\
&=1-\sum\limits_{i\in\{0,1\}}\hspace{0.1cm} \max_{L\in\cL_l}\hspace{0.12cm}\sum\limits_{x\in L} P_X(x)
W_o(i|x).\label{eq:list_priv_step3}
\end{align}
\noindent The proof of Theorem~\ref{thm:achiev} is completed upon showing that $\pi_{\rho}^{(l)}\left(W_o\right)$ above 
equals the right-side of~\eqref{eq:list-priv-binary} for $0\leq\rho\leq 1$ (and, in particular, the right-side 
of~\eqref{eq:list_priv_proof} for $0\leq\rho\leq\rho_1$). This is done in the following four steps.
\begin{enumerate}[1.]
    \item Using Lemma~\ref{lem:list_priv_ub}(i), (ii), we identify the set that attains the maximum 
    in~\eqref{eq:list_priv_step3} for $i$ in $\{0,1\}$.
    \item We first consider the case $\rho_1\leq\rho\leq 1$. Using the form of $\Lambda_{\rho}$ in~\eqref{eq:l_rho}, 
    we show that the set identified in Step 1 indeed attains the maximum in~\eqref{eq:list_priv_step3} for $i$ in $\{0,1\}$.
    \item Using Steps 1 and 2, for $\rho_1\leq\rho\leq 1$, we show that $\pi_{\rho}^{(l)}\left(W_o\right)=\pi_u^{(l)}(\rho)$ 
    in~\eqref{eq:list-priv-binary}.
    \item For $0\leq\rho< \rho_1$, using Steps 1 and 2, we show directly that $\pi_{\rho}^{(l)}\left(W_o\right)=
    \pi_u^{(l)}(\rho)$ in~\eqref{eq:list-priv-binary}.
\end{enumerate}

{\it Step 1}: With $\Lambda_{\rho}$ as in~\eqref{eq:l_rho}, and noting by Lemma~\ref{lem:list_priv_ub}(ii) that 
$l-|\Lambda_{\rho}|\leq |f^{-1}(i) \setminus \Lambda_{\rho}|, \ i\in\{0,1\}$, we claim that the maxima in~\eqref{eq:list_priv_step3} are 
attained by $L^i, \ i=0,1$, in $\cL_l$ given by 
\begin{equation}
    \label{eq:claim}
L^i = \Lambda_{\rho} \cup \left[f^{-1}(i) \setminus \Lambda_{\rho}\right]_{l-\left\lvert \Lambda_{\rho}\right\rvert}, \ \ i=0,1.
\end{equation} 
\noindent For any $A\subseteq\cX$, let $\left[A\right]^i$ denote the (not necessarily unique) $i$th-largest $P_X$-probability 
element in $A$, $i=1,\ldots,|A|$. Let $\left\lvert \Lambda_{\rho}\cap f^{-1}(i)\right\rvert=a_i, \ i\in\{0,1\}$. 
Using Lemma~\ref{lem:list_priv_ub}(i) in~\eqref{eq:claim}, said claim means that
\begin{align}
L^1&=\left[f^{-1}(1) \right]_{a_1+l-\left\lvert \Lambda_{\rho}\right\rvert}\cup\left[f^{-1}(0)\right]_{a_0}\nonumber\\
L^0&=\left[f^{-1}(1)\right]_{a_1}\cup \left[f^{-1}(0) \right]_{a_0+l-\left\lvert \Lambda_{\rho}\right\rvert} \nonumber
\end{align}
attain the maxima in~\eqref{eq:list_priv_step3}. 

{\it Step 2}: When $\rho_1\leq\rho\leq 1$, with the choice of $W_o$ in~\eqref{eq:achiev-scheme}, for the 
maximum in~\eqref{eq:list_priv_step3} for $i=1$ to be attained by $L^1$, it suffices if
\begin{equation}
\label{eq:claim1}
\rho P_X\left([f^{-1}(1)]^{a_1+l-|\Lambda_{\rho}|}\right)\geq
(1-\rho)P_X\left([f^{-1}(0)]^{a_0+1}\right)\end{equation}
\noindent and
\begin{equation}
\label{eq:claim2}
(1-\rho) P_X\left([f^{-1}(0)]^{a_0}\right)\geq
\rho P_X\left([f^{-1}(1)]^{a_1+l-|\Lambda_{\rho}|+1}\right).
\end{equation}
\noindent If $a_0=0$ or $a_1+l-|\Lambda_{\rho}|=|f^{-1}(1)|$, then it is enough to show~\eqref{eq:claim1}, 
as~\eqref{eq:claim2} is vacuous. If $a_0=|f^{-1}(0)|$ or $a_1+l-|\Lambda_{\rho}|=0$, then again it is adequate 
to show~\eqref{eq:claim2}, as~\eqref{eq:claim1} is vacuous. Since  $\Lambda_{\rho}$ attains the 
maximum for the expression within $[\cdot]$ in~\eqref{eq:l_rho},

\begin{multline}
\label{eq:max_eqn}
 P_X(\Lambda_{\rho}) + \rho\sum\limits_{i\in\{0,1\}} P_X\left(\left[f^{-1}(i)\setminus \Lambda_{\rho}\right]_{l-|\Lambda_{\rho}|} \right) \\ \geq 
 P_X(\Lambda) + \rho\sum\limits_{i\in\{0,1\}} P_X\left(\left[f^{-1}(i)\setminus 
 \Lambda\right]_{\min\{l-|\Lambda|,|f^{-1}(i)\setminus \Lambda|\}} \right), \ \ \Lambda \subset \cX, \ 0\leq |\Lambda|\leq l. 
\end{multline}
By choosing $\Lambda=\Lambda_{\rho}\cup[f^{-1}(0)]^{a_0+1}$ in\footnote{We know that since $\rho>\rho_1$, 
$|\Lambda_{\rho}|\leq l-1$, and therefore, $\Lambda=\Lambda_{\rho}\cup[f^{-1}(0)]^{a_0+1}$ is a valid choice 
for $\Lambda$ in~\eqref{eq:max_eqn}. Also, by a straightforward calculation, 
$l-|\Lambda|\leq |f^{-1}(i)\setminus \Lambda|, \ i=0,1.$}~\eqref{eq:max_eqn}, we get
\begin{multline*}
  P_X\left(\Lambda_{\rho}\right) + \rho\sum\limits_{i\in\{0,1\}} P_X\left(\left[f^{-1}(i)\setminus \Lambda_{\rho}\right]_{l-|\Lambda_{\rho}|} \right) \\ 
  \geq 
 P_X\left(\Lambda_{\rho}\cup[f^{-1}(0)]^{a_0+1}\right) + \rho\sum\limits_{i\in\{0,1\}} 
 P_X\left(\left[f^{-1}(i)\setminus \left(\Lambda_{\rho}\cup[f^{-1}(0)]^{a_0+1}\right)\right]_{l-|\Lambda_{\rho}|-1} \right),  
\end{multline*}
which reduces to~\eqref{eq:claim1}. Similarly, by choosing $\Lambda=
\Lambda_{\rho}\setminus[f^{-1}(0)]^{a_0}$ in\footnote{As noted, we only need to show~\eqref{eq:claim2} 
if $a_0>0$ and $a_1+l-|\Lambda_{\rho}|<|f^{-1}(1)|$. Then with $\Lambda=\Lambda_{\rho}\setminus[f^{-1}(0)]^{a_0}$, 
it can be verified easily that if $a_0>0$ and $a_1+l-|\Lambda_{\rho}|<|f^{-1}(1)|$, then 
$l-|\Lambda|\leq |f^{-1}(i)\setminus \Lambda|, \ i=0,1$.}~\eqref{eq:max_eqn}, we get~\eqref{eq:claim2}.

For establishing that the maximum in~\eqref{eq:list_priv_step3} for $i=0$ is attained by $L^0$, similarly it is sufficient to show that
\begin{equation}
\label{eq:rmnxjs}
\rho P_X\left([f^{-1}(0)]^{a_0+l-|\Lambda_{\rho}|}\right)\geq
(1-\rho)P_X\left([f^{-1}(1)]^{a_1+1}\right)\end{equation}
\noindent and
\begin{equation}
\label{eq:lcnaka}
(1-\rho) P_X\left([f^{-1}(1)]^{a_1}\right)\geq
\rho P_X\left([f^{-1}(0)]^{a_0+l-|\Lambda_{\rho}|+1}\right).
\end{equation}
This is done by choosing $\Lambda=\Lambda_{\rho}\cup[f^{-1}(1)]^{a_1+1}$ and $\Lambda=
\Lambda_{\rho}\setminus[f^{-1}(1)]^{a_1}$ which give~\eqref{eq:rmnxjs} and~\eqref{eq:lcnaka}, 
respectively, by similar calculations as for the case $i=1$.

{\it Step 3}: Finally, upon substituting~\eqref{eq:claim} in~\eqref{eq:list_priv_step3}, the resulting 
$\pi_{\rho}^{(l)}\left(W_o\right)$ equals the upper bound $\pi_{u}^{(l)}\left(\rho\right)$ in~\eqref{eq:list-priv-binary}. 
This completes the proof of achievability when $\rho_1\leq\rho\leq 1$.

{\it Step 4}: We now consider the case $0\leq\rho<\rho_1$. Having established at $\rho=\rho_1$ that 
$\pi_{\rho_1}^{(l)}\left(W_o\right)=\pi_u^{(l)}(\rho_1)$, and using~\eqref{eq:list_priv_proof}, we get 
$\pi_{\rho_1}^{(l)}\left(W_o\right)=\pi_u^{(l)}(\rho_1)=1-P_X(L_l^*)$. Then, recalling~\eqref{eq:achiev-scheme}, for $0\leq\rho<\rho_1$,
\[
\pi_{\rho}^{(l)}\left(W_o\right)=\pi_{\rho_1}^{(l)}\left(W_o\right)=1-P_X(L_l^*)=\pi_u^{(l)}(\rho),
\]
where the last equality is by~\eqref{eq:list_priv_proof}. This completes the proof of Theorem~\ref{thm:achiev}.
\qeed

\vspace{0.2cm}

\noindent{\it Remark}: Observe in the proof of Theorem~\ref{thm:list_priv_ub}(iii) that in obtaining the 
bound~\eqref{eq:bound_slack}, a list of size less than $l$ will be chosen for the second summation in the 
right-side of~\eqref{eq:sec-summ} if $\Lambda_{\rho}$ is such that, for some $i$ in $\cZ$, it holds that 
$l-|\Lambda_{\rho}|>|f^{-1}(i)\setminus\Lambda_{\rho}|$. This would suggest an apparent slackness in the 
upper bound on $\pi(\rho)$. We provide next an example to show that this is not necessarily the case. 
Consider the mapping $f:\{0,1,\ldots,4\}\rightarrow\{0,1,2\}$ with 
\begin{gather*}
    f(0)=f(1)=0, \ \ \ \  f(2)=f(3)=1 \ \ \ \ \text{and}  \ \ \ \ f(4)=2. 
\end{gather*}

\noindent The list size $l=2$, and $P_X$ is the uniform pmf on $\cX$. By a straightforward  calculation, we obtain
\[
\pi_u^{(l)}(\rho)=\begin{cases}
    1-\rho, \ \ \ &\frac{1}{2}\leq\rho\leq 1\\
        \frac{4}{5}-\frac{3}{5}\rho, \ \ \ &\frac{1}{3}\leq\rho\leq \frac{1}{2}\\
            \frac{3}{5}, \ \ \ &0\leq\rho\leq \frac{1}{3}.
\end{cases}
\]
\noindent Observe that for $1/2\leq\rho\leq 1$, $\Lambda_{\rho}=\emptyset$, due to which 
$l-|\Lambda_{\rho}|=2>|f^{-1}(2)\setminus \Lambda_{\rho}|=1$. But the $\rho$-QR $W:\{0,1,\ldots,4\}\rightarrow\{0,1,2\}$ given by
\[W(i|x)=
\begin{cases}
    \rho, \ \ \ &i\in\{0,1,2\}, \ \ \ x\in f^{-1}(i)\\
    1-\rho, \ \ \ &i=0, \ \ \ x\in \{2,3\}\\
    1-\rho, \ \ \ &i=1, \ \ \ x\in \{0,1\}\\
    \frac{1-\rho}{2}, \ \ \ &i\in\{0,1\}, \ \ x=4\\
    0,\ \ \ &\text{otherwise},
\end{cases}\]
\noindent has list privacy $1-\rho$ for $1/2\leq\rho\leq 1$. Therefore, for $1/2\leq\rho\leq 1$, 
$\pi_u^{(l)}(\rho)$ is optimal even though $l-|\Lambda_{\rho}|=2>|f^{-1}(2)\setminus \Lambda_{\rho}|=1$.

\section{Discussion}
\label{sec:list-priv-disc}
Theorem~\ref{thm:list_priv_ub} provides a general upper bound for list $\rho$-privacy that is tight for the special case of a 
binary-valued function. What is the exact characterization of list $\rho$-privacy for functions that are not binary-valued? With 
regard to the converse, from the derivation of $\pi_u^{(l)}(\rho)$, we get that the querier's best list estimate, based on $i$ in $\cZ$, 
has the structure that it consists of elements from the highest $P_X$-probability elements in $\cX$ and the highest $P_X$-probability 
elements in $f^{-1}(i)$. This natural form of the querier list estimate suggests the optimality of $\pi_u^{(l)}(\rho)$. Hence, it 
promotes the conjecture that list $\rho$-privacy $\pi^{(l)}(\rho)$ equals $\pi_u^{(l)}(\rho)$. If true, what is the $\rho$-QR that 
attains a list privacy of $\pi_u^{(l)}(\rho)$ for functions that are not binary-valued? Our optimal $\rho$-QR $W_o$~\eqref{eq:achiev-scheme} 
for a binary-valued function involves only $\rho_1$. Recalling $\rho_1,\ldots,\rho_k$ from Lemma~\ref{lem:list_priv_ub}(v), 
what role, if any, will $\rho_2,\ldots,\rho_k$ play in the construction of the optimal $\rho$-QR for functions that are not 
binary-valued? Further, we note in~\eqref{eq:achiev-scheme} that $W_o$ corresponding to $f^{-1}(0)$ (resp. $f^{-1}(1)$) are 
identical. Such a $W_o$ has the attribute that the resulting $\rho$-QR $F(X)$ satisfies the Markov property $X\MC f(X)\MC F(X)$, 
and is termed an ``add-noise'' $\rho$-QR in{\cite[Definition $2$]{Nages19}}. Will $\rho$-QRs, not necessarily of the 
add-noise type, be required for achieving list $\rho$-privacy in general? These questions are open.

\appendices
\section{Proof of Lemma~\ref{lem:list_priv_ub}}
\label{app:lemma-pf}
\vspace{0.1cm}
\noindent (i) We need to show that 
\begin{equation}
\label{eq:property}
\Lambda_{\rho}\cap f^{-1}(i)=\left[f^{-1}(i)\right]_{\left\lvert \Lambda_{\rho}\cap f^{-1}(i)\right\rvert}, 
\ i\in\cZ.
\end{equation}

Writing the term within $[\cdot]$ in~\eqref{eq:l_rho} with $\Lambda=\Lambda_{\rho}$ as
\begin{equation}
\label{eq:3.8} 
\sum_{i\in\cZ} \left[ P_X\left(\underbrace{\Lambda_{\rho}\cap f^{-1}(i)}\right) + \rho 
P_X\left(\underbrace{\underbrace{\left[f^{-1}(i)\setminus \Lambda_{\rho}\cap f^{-1}(i)\right]_{\min\{l-|\Lambda_{\rho}|,|f^{-1}(i)\setminus 
\Lambda_{\rho}|\}}}}\right)\right].
\end{equation}
In~\eqref{eq:3.8}, for each $i\in\cZ$, focusing on the term within $[\cdot]$, if~\eqref{eq:property} is violated, there exists an 
element in $\underbrace{\underbrace{\cdot}}$ of higher 
$P_X$-probability than the minimum 
$P_X$-probability element in $\underbrace{\cdot}$. Swapping them would lead to a higher value of $[\cdot]$ since $\rho\leq 1$, 
thereby contradicting the optimality of $\Lambda_{\rho}$ in~\eqref{eq:l_rho}. Therefore,~\eqref{eq:property} holds. 
\vspace{0.1cm}

\noindent (ii) Assume 
\begin{equation}
    \label{eq:eqn-contradiction}
l-|\Lambda_{\rho}|> |f^{-1}(0)\setminus \Lambda_{\rho}|
\end{equation}
and
\[
l-|\Lambda_{\rho}|\leq |f^{-1}(1)\setminus \Lambda_{\rho}|.
\]
\noindent Let
\[
\label{eq:Li-defn}
\left\lvert \Lambda_{\rho}\cap f^{-1}(i)\right\rvert =a_i, \ \ \ i\in\{0,1\},
\]
which give
\begin{equation}\label{eq:rewweer}
|\Lambda_{\rho}|=\sum_{i\in\{0,1\}} \ a_i\end{equation}
\noindent and
\begin{equation}\label{eq:papcapda}
    |f^{-1}(i)\setminus \Lambda_{\rho}|=|f^{-1}(i)|-a_i,  \ \ \ i\in\{0,1\},
\end{equation}
and using Lemma~\ref{lem:list_priv_ub}(i),
\begin{equation}
\label{eq:Lrho-exact-defn}
    \Lambda_{\rho}=\left[f^{-1}(1)\right]_{a_1}\cup \left[f^{-1}(0)\right]_{a_0}.
\end{equation}
Substituting~\eqref{eq:rewweer},~\eqref{eq:papcapda} in~\eqref{eq:eqn-contradiction}, we get
\begin{equation}
    l-|f^{-1}(0)| >  a_1 \label{eq:mammmmm1}.
\end{equation}
Consider the set $\tilde{\Lambda}_{\rho}$ given by
\begin{equation}
\tilde{\Lambda}_{\rho}=\left[f^{-1}(1)\right]_{l-|f^{-1}(0)|}\cup \left[f^{-1}(0)\right]_{a_0}. \label{eq:alt-set}
\end{equation}
\noindent 
Since $l-|f^{-1}(0)|>a_1\geq 0$, $|f^{-1}(1)|>l-|f^{-1}(0)|$ and $\left\vert\tilde{\Lambda}_{\rho}\right\vert=l-|f^{-1}(0)|+a_0\leq l$,  
$\Lambda=\tilde{\Lambda}_{\rho}$ is a valid choice for the expression within $[\cdot]$ in~\eqref{eq:l_rho}. We claim that 
choosing $\Lambda=\tilde{\Lambda}_{\rho}$ gives a larger value for the expression within $[\cdot]$ in~\eqref{eq:l_rho}, 
thereby contradicting the optimality of $\Lambda_{\rho}$, and, in turn, proving that~\eqref{eq:eqn-contradiction} cannot hold. 
To establish the claim above, with $\Lambda=\Lambda_{\rho}$ in the expression within $[\cdot]$ in~\eqref{eq:l_rho}, and 
using~\eqref{eq:Lrho-exact-defn}, we get
\begin{align}
 [\cdot]= &P_X\left([f^{-1}(0)]_{a_0}\right) + \rho P_X\left(\left[f^{-1}(0)\setminus [f^{-1}(0)]_{a_0}\right]_{|f^{-1}(0)|-a_0} \right)+\nonumber\\
 & P_X\left([f^{-1}(1)]_{a_1}\right) + \rho P_X\left(\left[f^{-1}(1)\setminus [f^{-1}(1)]_{a_1}\right]_{l-a_1-a_0} \right);\label{eq:mcmakdkalk}
\end{align}
and upon choosing $\Lambda=\tilde{\Lambda}_{\rho}$ in $[\cdot]$ in~\eqref{eq:l_rho}, and using~\eqref{eq:alt-set}, we get
\begin{align}
 [\cdot]=& P_X\left([f^{-1}(0)]_{a_0}\right) + \rho P_X\left(\left[f^{-1}(0)\setminus [f^{-1}(0)]_{a_0}\right]_{|f^{-1}(0)|-a_0} \right)+\nonumber\\
 & P_X\left([f^{-1}(1)]_{l-|f^{-1}(0)|}\right) + \rho P_X\left(\left[f^{-1}(1)\setminus [f^{-1}(1)]_{l-|f^{-1}(0)|}\right]_{|f^{-1}(0)|-a_0} \right).
 \label{eq:vkadskcnslakc}
\end{align}
Observe that in~\eqref{eq:mcmakdkalk},~\eqref{eq:vkadskcnslakc}, by~\eqref{eq:mammmmm1}, 
$\left[f^{-1}(1)\right]_{a_1}\subset \left[f^{-1}(1)\right]_{l-|f^{-1}(0)|}$, and since $\rho\leq 1$, the claim holds. 

Similar arguments apply if we assume 
\[l-|\Lambda_{\rho}|>|f^{-1}(1)\setminus \Lambda_{\rho}| \ \  \text{and} \ \  l-|\Lambda_{\rho}|\leq|f^{-1}(0)\setminus \Lambda_{\rho}|\] 
\noindent or 
\[l-|\Lambda_{\rho}|>|f^{-1}(1)\setminus \Lambda_{\rho}| \ \  \text{and} \ \ l-|\Lambda_{\rho}|>|f^{-1}(0)\setminus 
\Lambda_{\rho}|.\] Therefore,~\eqref{eq:l-rho-not-empty} is true.

\vspace{.2cm}

\noindent (iii) Consider $
1-\pi^{(l)}_u(\rho) = \max\limits_{0\leq l'\leq l}\hspace{0.1cm} \alpha\left(\rho,l'\right)
$ where
\begin{gather*}
\alpha\left(\rho,l'\right) = \max_{\substack{\Lambda \subset \cX: \\ |\Lambda|=l'}} \hspace{0.1cm} \beta\left(\rho,l',\Lambda\right)\\
\beta\left(\rho,l',\Lambda\right)= P_X(\Lambda) + \rho\sum\limits_{i\in\cZ} P_X\left(\left[f^{-1}(i)\setminus \Lambda\right]_{\min\{l-l',\left\vert f^{-1}(i)
 \setminus \Lambda\right\vert\}}\right),
\end{gather*}
\noindent where $|\Lambda|=l'$. For each fixed $0\leq l'\leq l$ and fixed $\Lambda\subset \cX$ with 
$|\Lambda|=l'$, $\beta\left(\rho,l',\Lambda\right)$ 
is affine and nondecreasing in $\rho$. Hence, $\alpha\left(\rho,l'\right)$ -- as the maximum of 
finitely many affine and nondecreasing functions in $\rho$ -- is piecewise affine, continuous and 
nondecreasing in $\rho$. In turn, $1-\pi^{(l)}_u(\rho)$ -- as the maximum of $l+1$ such functions 
$\alpha\left(\rho,l'\right)$ -- also is piecewise affine, continuous and nondecreasing in $\rho$. The 
values of $\pi_u^{(l)}(\rho)$ at the end points $\rho=0$ and $\rho=1$ in~\eqref{eq:list_priv_extremes} 
follow from a straightforward calculation.\vspace{0.2cm}

\noindent(iv) For $l'=0$, $\alpha(\rho,0)$ is linear in $\rho$ with $\alpha(0,0)=0$ and 
$\alpha(1,0)=\sum\limits_{i\in\cZ}P_X\left(\left[f^{-1}(i)\right]_{\min\{l,\left\vert f^{-1}(i)\right\vert\}}\right)$; and 
for $l'=l$, $\alpha(\rho,l)\equiv P_X\left(L_l^*\right)$. And for $0<l'<l$, 
$\alpha\left(\rho,l'\right)$ is piecewise affine, continuous and nondecreasing in 
$\rho\in [0,1]$ as noted in the proof of (iii). Further, observe that for 
$0\leq l'_1<l'_2\leq l$, 
\[
\alpha\left(0,l'_1\right)\leq \alpha\left(0,l'_2\right) \ \text{while} \
\alpha\left(1,l'_1\right)\geq \alpha\left(1,l'_2\right)
\]
\noindent where the latter is immediate upon examining 
\[
\alpha\left(1,l'\right)= \max_{\substack{\Lambda \subset \cX: \\ |\Lambda|=l'}}
\left[P_X(\Lambda) + \sum\limits_{i\in\cZ} P_X\left(\left[f^{-1}(i)\setminus \Lambda\right]_{\min\{l-l',\left\vert f^{-1}(i)\setminus 
\Lambda\right\rvert\}}\right)\right]
\]
\noindent and observing that $
\alpha\left(1,l'\right)\leq \alpha\left(1,l'-1\right)$, etc. The claim in (iv) follows.\vspace{0.2cm}

\noindent (v) The claim follows readily from $|\Lambda_{\rho}|$ taking values in $\{0,1,\ldots,l\}$, property (iv), and $\rho$ ranging continuously in $[0,1]$.
\qeed

\section*{Acknowledgments}
We thank Himanshu Tyagi for a helpful discussion of data privacy~\cite{Nages19} that motivated the 
concept of list privacy. We also thank two anonymous referees for their constructive
suggestions which made for an improved presentation of our results.

\end{document}

%% file: preamble.tex
\newcommand{\cL}{{\mathcal L}}

\newcommand{\cX}{{\mathcal X}}

\newcommand{\cZ}{{\mathcal Z}}

\usepackage{color}

\theoremstyle{remark}
\newtheorem*{remark*}{Remark}
\newtheorem*{remarks*}{Remarks}